\documentstyle{article}
\begin{document}
\pagestyle{headings}
\noindent
\large
\begin{center}
Particle Path Formulation of Quantum Mechanics\\[8ex]
by\\[2ex]
S. R. Vatsya\\[6ex]
Centre for Research in Earth and Space Science\\[1ex]
York University\\[1ex]
North York, Ontario, Canada~~~~~M3J 1P3\\[2ex]
e-mail: VATSYA@Physics.UManitoba.CA           \\[5ex]
Presented in the conference\\[1ex]
The Present Status of the Quantum Theory of Light\\[1ex]
August 27 - 30, 1995\\[1ex]
York University,  Toronto, Canada\\[7ex]
\end{center}
%
%
\begin{abstract}
An extension of the classical action principle obtained in the framework of 
the gauge transformations, is used to describe the motion of a particle. 
This extension assigns many, but not all, paths to a particle. 
Properties of the particle paths are shown to impart wave like behaviour to 
a particle in motion and to imply various other assumptions and conjectures
attributed to the formalism of Quantum Mechanics. The Klein-Gordon and other
similar equations are derived by incorporating these properties in the
path-integral formalism. 
\end{abstract}
\section{Introduction}

This paper describes a recent approach to mechanics based on an extension of
Hamilton's action principle, obtained by a process of completion in
the framework of the gauge transformations.  

In Sec. 2, a motivation for the extension is developed by examining
the action principle and by reformulating it in terms of the gauge
transformations. In Sec. 3, the extension termed the gauge mechanical
principle, is formulated, interpreted and its solutions are
classified. In Sec. 4 the solutions of the gauge mechanical principle
are used to describe the motion of a free particle, the behaviour of
the particles in a double slit experiment and the Aharonov-Bohm
effect. Although the present formulation excludes some trajectories
from the collection of physical paths, the results are sufficient to
justify Feynman's path integral formalism to formulate mechanics, at
least approximately (Sec. 5). The properties of the physical paths are
therefore incorporated in the path-integral formalism in Sec. 5, to
derive a generalized Schr\"{o}dinger type equation which is then reduced
to a set of infinitely many four-dimensional equations, one of them
being the Klein-Gordon equation. In Sec. 6, some additional results
are quoted and directions for further development are indicated. In
conclusion in Sec. 7, summary of the results is used to justify the
present formulation of mechanics.

This approach to mechanics was developed independently of any direct 
considerations of the behaviour
of particles in experimental settings including the double slit
experiment. However, its implications lead one to consider the
following experimental observations and somewhat unorthodox conclusions 
that might be drawn from them.

In the double slit experiment, photons, electrons and other physical
entities that are normally considered particles, demonstrate their
particle nature if observed individually. However, if many are allowed
to pass through the slits, together or one after the other, then an
interference-like pattern of intensity emerges on the screen [1].
Since a wave would produce such a pattern, it is assumed that each
particle also has a wave character. Quantum Mechanics accepts this
duality by attaching a probability wave with a particle in motion i.e.
the wave determines the probability of finding a particle in a certain
space-time region. This fusion of wave and particle nature creates
most of the logical difficulties with Quantum Mechanics [1,2]. The
observation in the double slit experiment is viewed about the most
puzzling mystery of nature. Also, its understanding is considered
pivotal to the resolution of most of the paradoxical situations
arising in the microscopic phenomena described by Quantum Mechanics.

While a wave would produce the intensity distribution of the type
observed in the double slit experiment, the converse is not
necessarily true i.e. the observation of this pattern does not prove
that it was produced by a wave. A closer scrutiny of the experimental
observations suggests an alternative possibility.

Conclusions based on relevant observations identify the observed
entity as a particle when emitted or absorbed.
Experiments designed to reveal its wave nature during travel observe each
individual with particle like attributes. 
Therefore it may be possible to describe the experimental
observations by associating a particle like trajectory with each of
the entities. These observations on a number of particles suggest the
possibility of the existence of a collection of paths out of which
each particle takes one, probably randomly. This collection must be
endowed with some characteristics which are responsible for the
inclusion of more paths ending about the bright regions and exclusion
of others. Therefore it appears more reasonable to build a theory of
mechanics by characterizing the collection of particle paths rather
than attempting to fuse mutually exclusive wave and particle
behaviours. If this view is adopted, then the effect of an observation
on its outcome must be the result of the disturbance suffered by the
particle and hence, must be described in this manner. This philosophy
has its origin in Fermat's principle of stationary time in light and
Hamilton's principle of stationary action in classical mechanics. Both
of these theories are geometrical in nature instead of mechanical,
although Hamilton's principle is equivalent to Newton's second law
which gives an impression of being a mechanical theory.

The implications of the present extension of Hamilton's action principle 
are in accordance with inferences that could be drawn from the experimental 
observations as discussed above.  To be precise, the extension yields 
a collection of infinitely many, but not all, paths for a particle to follow 
which are endowed with some properties
by virtue of the fact that they are the solutions of the extended
principle. These properties are shown to describe the behaviour of
particles in a double slit experiment and in the Aharonov-Bohm
experiment without invoking the usual assumption of probability waves
or the formalism of Quantum Mechanics. The results are shown to
justify Feynman's path integral formulation and used in this framework
to derive a generalized Schr\"{o}dinger type equation. Properties of the
particle paths are used to reduce the general equation into infinitely
many four dimensional equations, one of them being the Klein-Gordon
equation.

This formulation yields as results, the assumptions underlying the
standard Quantum Mechanics and various other intuitive conjectures
usually attributed to the formalism of Quantum Mechanics. However
there are some differences between the consequences of the present
formulation and the standard Quantum Mechanics which are indicated in
the sequel.

This paper is more detailed than the paper
to appear in the conference proceedings.

\section{The Action Principle}

In this section, a reformulation of the
action principle is presented that is well suited for its extension in
the framework of the gauge transformations.

Let $L(\dot{x},x,\tau)$ be a Lagrangian defined on curves in a manifold
${\cal M}$. While the action principle may be formulated in any
differentiable manifold, for the present we shall have occasion only
to deal with the Minkowski space. For a path $\rho(AB) = x(\tau)$ 
with $x(\tau_1)=A$, $x(\tau_2)=B$, the action functional 
${S}_{BA}(\rho) = S(\tau_1,\tau_2)$ is given by 
\begin{equation}
   S(\tau_1,\tau_2) = \int_{\tau_1}^{\tau_2} L(\dot{x},x,\tau) d\tau
\end{equation}
The action principle characterizes the particle path(s) by requiring
the action to be stationary i.e.
\begin{equation}
\delta {S} = {S}_{BA}(\rho') - {S}_{BA}(\rho) = 0
\end{equation}
up to the first order in $\delta x$ where $(x+\delta x) (\tau) = \rho'$. The
end points A and B are kept fixed and correspond to the same parameter
values as the undistorted curve i.e. $\tau_1$ and $\tau_2$ respectively.
Eq. (2) is expected to hold for all curves $\rho'$ in a small
neighbourhood of the solution $\rho$ if it exists.

Some conceptual clarity is gained in describing the action principle
by considering the analogue of $x(\tau)$ in ${\cal M'}$ obtained from
${\cal M}$ by including $\tau$ as an additional coordinate [3, Ch. 1.1]. 
Thus the curve $x(\tau)$ in ${\cal M}$ corresponds to the set of points
$(x(\tau),\tau)$ in ${\cal M'}$, eliminating a need for an explicit reference
to the parameterization. 

Eq. (2) in ${\cal M'}$ takes the following form:
\begin{equation}
S_{ABA}(\rho_{\!_c}) = 0
\end{equation}
up to the first order in $d\sigma$ where $\rho_{\!_c}$ is the closed curve
in ${\cal M'}$ obtained as the union of $\rho'$ and $\rho$ inverse i.e. 
$\rho_{\!_c}$ traces the path $\rho'$ from $(A,\tau_1)$ to $(B,\tau_2)$ 
and then inverted $\rho$ from $(B,\tau_2)$ to $(A,\tau_1)$, and $d\sigma$ 
is the area enclosed by $\rho_{\!_c}$. Eq. (2), equivalently (3) yields 
the Euler-Lagrange equation that describes the particle path.

Consider a charged particle in an electro-magnetic field which may be
described by the Lagrangian $L = L^P - \phi_\mu \dot{x}^\mu$, where 
$L^P = \frac{1}{2} m (\dot{x}_\mu\dot{x}^\mu + 1)$, 
and $\phi_\mu$ are the electro-magnetic potentials. 
Quite frequently, a homogeneous Lagrangian is
used instead, but $L$ is more convenient. Both formulations are
equivalent with $\tau$ being the proper time. The Lorentz equation
describing the path of a charged particle in an electro-magnetic field
is the solution of (2) or (3) with the Lagrangian given by $L$, i.e.
the particle path is characterized by
\pagebreak[1]
{\samepage
\begin{equation}
 \oint L^P d\tau - \oint \phi_\mu dx^\mu = 0
\end{equation}
up to the first order.
}

Eq. (4) relates this characterization of the particle-path with the
gauge transformations as follows. Weyl introduced the notion of the
gauge transformations by proposing that a rigid measuring rod must be
gauged at every space-time point according to the rule [4]
\begin{equation}
  d\Phi_A = \alpha \phi_\mu dx^\mu \Phi_A
\end{equation}
where $d\Phi_A$ is the change suffered by a rod of length  $\Phi_A$
at $A$ under the infinitesimal displacement $dx$ and $\alpha$ is a constant. 
From (5), the length 
$\Phi_{BA}$ at $B$ of the same rod transported along $\rho$ is given by
\begin{equation}
    \Phi_{BA}(\rho) = U_{BA}(\rho)\Phi_A
\end{equation}
where 
\begin{equation}
  U_{BA}(\rho) = {\rm Exp} \left( \alpha \int_{\rho(AB)} 
                              \phi_\mu(x)dx^\mu \right) 
\end{equation}
is the group element associated with $\rho(AB)$. The Lie algebra element
associated with the displacement $dx$ is  $\alpha \phi_\mu dx^\mu$.
It is clear that the action principle describes the particle-path in
terms of the gauge Lie algebra element, associated with the curves 
of the type $\rho_c$.

Eq. (3) may be expressed as 
\begin{equation}
{\rm exp}(\alpha {S}_{ABA}(\rho_{\!_c})) = 1
\end{equation}
equivalently as
\begin{equation}
{U^P}_{ABA}(\rho_{\!_c}) = U_{ABA}(\rho_{\!_c}) 
\end{equation}
up to the first order in $d\sigma$, with 
\( {U^P}_{BA}(\rho) = {\rm exp}(\alpha {S^P}_{BA}(\rho)) \)
where ${S^P}_{BA}(\rho)$ is the free particle part of the action associated with
$\rho(AB)$.

\section{The Gauge Mechanical Principle}  

The formulation of the action
principle described in Sec. 2 indicates that the classical description
of motion is deficient in gauge group theoretical terms. This
description limits itself to a characterization of particle-path(s) in
terms of the Lie algebra elements, equivalently, the infinitesimal
gauge group elements, which is accurate only up to the first order.
This characterization is local in nature. Additional information that may
be available in the global group elements is not utilized in the
action principle. Therefore a description in terms of the group
elements should be expected to be more complete. This deficiency can
easily be corrected by including the higher order terms in addition to
other adjustments if need be. Any such characterization must reduce to 
\begin{equation}
{\rm exp}(\alpha {S}_{ABA}(\rho_{\!_c})) = 1
\end{equation}
Eq. (10), although an extension, limits itself to considering only the closed
curves in ${\cal M'}$ while the group elements are defined for all
curves. If the action principle is to be extended in terms of the
gauge group elements, then this restriction becomes redundant. To
achieve appropriate generality consistent with the domain of
definition of the gauge group elements, the action principle should be
extended to
\begin{equation}
\kappa^{-1}(B) {\rm exp}(\alpha {S}_{BA}(\rho)) \kappa(A) = 1 
\end{equation}
where $\kappa$ is as yet an undetermined function which cancels out for the
closed curves. The characterization of particle paths by (11) has been
termed the gauge mechanical principle [5]. Its solutions will be
called the physical paths which a particle is allowed to follow.

It should be remarked that there is no logical deficiency or
inconsistency in the action principle itself. The argument here is
that the action principle provides an incomplete description of motion
in gauge group theoretical terms. Prejudice in favour of the group
elements in comparison with the Lie algebra elements, in favour of the
global in comparison with the local, is a matter of metaphysical
conviction.

In the above, we have provided arguments to justify the present
extension of the classical action principle, not a derivation of the
gauge mechanical principle. These arguments are to some extent
irrelevant as far as the matter of the extension is concerned. The
fact that (11) reduces to (8) with appropriate restrictions is
sufficient to prove that (11) is an extension of the action principle.

Furthermore, the gauge mechanical principle by itself may be made the
basis of a formulation of mechanics whether it is an extension of the
action principle or not. All that is required is that it provide an
adequate description of the motion of particles. The fact that it is
an extension of the action principle serves only to relate the
resulting mechanics with Classical Mechanics. In the remainder of this
section we clarify the principle further and present its alternative
statements. 

First we relate the gauge mechanical principle with Newton's second
law of motion. Some such relation should be expected as the action
principle is equivalent to Newton's law. We limit here to the motion
of a charged particle in an electro-magnetic field which illustrates
the relation without cluttering the concepts with unnecessary
generalities.

Eq. (11) may be expressed as 
\begin{equation}
{U^P}_{BA}(\rho) = \kappa(B) U_{BA}(\rho) \kappa^{-1}(A)
\end{equation}
The right side of (12) is equal to $(1 + \alpha F_{\mu\nu} d \sigma^{\mu\nu})$ 
for infinitesimal closed curves, where $F_{\mu\nu}$ are the components
of the field tensor. The left side under the same conditions reduces
to $(1-(p_\mu \dot{p}_\nu - \dot{p}_\mu p_\nu) d \sigma^{\mu\nu})$
where $p_\mu$ are the components of the canonical momentum. This
equality is equivalent to the Lorentz equation, equivalently, Newton's
second law [6].

The gauge mechanical principle may also be interpreted in terms of
Weyl's original notion of gauging a rigid measuring rod as follows.
Recall that $\Phi_{BA}$ is the length of Weyl's rod at B transported along
$\rho(AB)$ while its length at A was $\Phi_A$. Eq. (12) may be expressed as 
\begin{equation}
  \Phi_{BA}^P(\rho) = \kappa(B) \Phi_{BA}(\rho)
\end{equation}
where $\Phi_{BA}^P(\rho) = U_{BA}^P(\rho)\Phi_A^P$, and 
$\Phi_{A}^P = \kappa(A) \Phi_A$.
Weyl's gauge transformations determine the effect of a field on the
rigid measuring rod. One may take another rod of length $\Phi_A^P$ 
and transport it along a given curve $\rho(AB)$. Let $\Phi_{BA}^P$
be its length at $B$ determined as above without any reference to the
field. The gauge mechanical principle requires that Weyl's gauge and
the present gauge must return essentially in the same relation as they
began with at $A$ for $\rho(AB)$ to be a particle path. 

In Newton's second law, one equates a force-like quantity determined
solely by the curves in the space-time manifold with the force
postulated by an independent law. In the present formulation, one
computes the change in the length of the measuring rod solely from the
curves in ${\cal M'}$ without any reference to the field, which is then
related to the change in Weyl's rod. It is not necessary to set 
$\Phi_A^P =\Phi_A$ as it would limit generality without adequate justification. 
It is sufficient that a precise map between $\Phi_A^P$, $\Phi_{BA}^P$ 
and $\Phi_A$, $\Phi_{BA}$ be available. This is consistent with (11) and (12), 
as the equality (11) for closed curves implies only the group
equivalence (12) for general curves. 

The function $\kappa$ in the above appears as a requirement of the
mathematical generality as there is no justification for imposing further
restrictions on (12). However, for a physical theory, $\kappa$ must have a
clearer physical significance which we discuss below. 

The elements $U$ and $U^P$ appearing in (12) pertain to the interiors
of the respective curves. As such there is no consideration of the
initial physical state of a particle or of local interventions at B or
elsewhere. Obviously the physical paths for two particles in different
physical states should be expected to be different in the same field.
Therefore it is legitimate to interpret $\kappa$ as representing the
physical state of the particle. Interaction with the detecting
instrument is local in nature and has a direct impact on the physical
state of the particle. Therefore such interactions are also included
in $\kappa$ by way of the physical state of the particle. A precise
computation of $\kappa$ is not necessary for a variety of experimental
situations. For example, in the double slit experiment, the particles
passing through two slits at A and A' are prepared by the same
physical process and are identical in every other respect. Therefore,
it is legitimate to conclude that particles at A and A' are in the same 
physical state even if it may not be precisely defined.  Hence, we may set 
$\kappa(A)=\kappa(A')$. Similarly, two beams meeting at B interact with
the same instrument. Therefore B is not only geometrically the same
point for two paths $\rho(AB)$ and $\rho(A'B)$, it is also physically
equivalent. Therefore $\kappa$ has the same value for two beams at B.
This will be found sufficient for the description of the behaviour of
particles in the double slit experiment. The same comment applies to
various other experimental situations.

Consider a free particle travelling from A to B along $\rho(AB)$ without
any interactions including the intrusion of a detecting instrument. In
this situation, the physical state of the particle must remain
unchanged. Therefore we shall assume that for a free particle
$\kappa(A)=\kappa(B)$ for all points A and B i.e. $\kappa$ is constant. This
extends Newton's first law. The effects of interactions on $\kappa$ may also be
computed. A detailed description of such computations is beyond the scope of
this article but it is indicated below to an extent necessary for the clarity
of the gauge mechanical principle. 

Since for a free particle $\kappa$ is constant, any change in its value
must be a result of an interaction. In standard interventions, a
precise value of the interaction is unknown e.g. a detecting
instrument but the instantaneous change in the classical momentum may be
computed or estimated with sufficient accuracy. This information is
sufficient to compute the change $\Delta S$ in the action caused by the
interaction. The change in the value of $\kappa$ is then given by 
${\rm exp}(\alpha \Delta S_{BA} (\rho))$.

The value of $\alpha$ still remains undetermined which we obtain below.
For a free particle, the physical paths are defined by
\begin{equation}
{\rm exp}(\alpha {S}_{BA}(\rho)) = 1
\end{equation}
In general, the action $S_{BA}(\rho)$ is real and non-zero. There are
configurations of curves with total action equal to zero but there is
no justification for restricting the description of motion to such
curves only. Therefore $\alpha$  must be purely imaginary which may be set
equal to $i$ in appropriate units.

The representations of the gauge mechanical principle given by (11),
(12) and (13) are essentially equivalent. Reference to the gauging of
the measuring rod is inconsequential for the following developments.
Reference to one of the representations, therefore, will include others
as well.

Solutions of (13) are identified by their equivalence classes as
follows.  
Let $V_{BA}(\rho) = \kappa(B) V_{BA}'(\rho) \kappa^{-1}(A)$ with 
$V_{BA}'(\rho) = U_{AB}^P(\rho)U_{BA}(\rho)$,
and let $\{B_j\}$ be a set of points on $\rho(AB)$ 
such that $V_{B_jA}(\rho)\Phi_A^P = \Phi_A^P$.  If one member
of $\{B_j\}$ is a physical point with respect to $\{ \rho(AB),\Phi_A^P \}$, 
then this is also the case
for each $j$.  Thus the equivalence class $\{B_j\}$ so defined 
characterizes the
solutions $\{\rho(AB_j)\}$. A natural order is defined on $\{B_j\}$ 
by setting  $B_j$ to be the $j$th
closest member to $A$.  Let $\{B_j^k\}$, $k=1,2,....$, 
be such ordered equivalence classes with respect to 
$\{\rho(AB^k),\Phi_A^P\}$.
The set $\zeta_j = \{B_j^k \}$  defines a physical `surface' for each $j$.

For a free particle, the physical paths are the 
solutions of (14) which reduces to
\begin{equation}
\exp \left( im \int_{\rho(AB)} u_\mu dx^\mu \right) \Phi_A^P = \Phi_A^P
\end{equation}
The equivalent points
$\{B_j\}$ on these curves satisfy
\[  m \int_{\rho(B_j,B_{j+1})} u_\mu dx^\mu = 2 \pi    \]
Along the paths characterized by a constant velocity $\bar{u}$, $B_j$ and 
$B_{j+1}$ are thus separated by the de
Broglie wavelength $2 \pi / m \bar{u}$ and the length of a physical path
is its integral multiple.

Consider a source-detector system with source at $A$ and detector at $B$. 
A curve $\rho (AB)$ will be called monotonic if the parameter value increases
or decreases monotonically along the curve.  By convention, $\tau$ will be
assumed to increase from $A$ to $B$.  A particle starting at $A$ and confined
to $\rho (AB)$ is observable at $B$ if and only if $\rho (AB)$ is physical.
If $\theta$  is the intensity associated with  
$\rho (AB)$ at $A$ then the intensity transmitted
to $B$ by this path must be equal to $\theta$.

A union of physical paths is obviously physical.  Also a union of
non-physical monotonic curves can be physical.  For example, let $\rho (AB)$
be a monotonic physical path with the associated physical points $\{ B_j \}$
and let $C$ be a point in the interior of $\rho(B_j B_{j+1})$. 
Then the union of $\rho (B_j C)$ and $\rho (CB_{j+1})$ is 
$\rho (B_{j} B_{j+1})$ 
and the union of $\rho(B_{j-1} C)$ and $\rho (CB_j)$ is 
$\rho (B_{j-1} B_{j})$,
both of which are physical.  However, these trivial constructions are
redundant as they are indistinguishable from the paths of the type 
$\rho(B_k B_{k+l})$. 
A significant, non-trivial class of such paths is described below.

Consider a configuration of two curves  
$\rho(AB)$ and  $\rho'(AB)$ with  $\rho_{\!_c}(ABA)$
being the union of $\rho'(AB)$ and $\rho(BA)$.
According to the
present prescription, if (13) is satisfied then this is a physical
configuration.  Since the evolution parameter increases from $A$ to $B$ along
both of the curves, particle must travel from $A$ to $B$ along $\rho(AB)$ and
$\rho'(AB)$. Therefore $\rho(AB)$ and $\rho'(AB)$ offer equally likely
alternatives for the transmission of a particle from $A$ to $B$, even if 
$\rho(AB)$ and $\rho'(AB)$ may not be physical.  
The case of the alternatives of
the type $\rho(AB)$ and $\rho'(CB)$ is treated similarly.  
To be precise, let the parameter value at B be $\tau_{\!_B}$.  
According to the above convention, $\tau$
increases from $C$ to $B$ along $\rho'(CB)$ and decreases from 
$B$ to $A$ along $\rho(BA)$.  
If $V_{CBA} (\rho'') \Phi_A^P = \Phi_A^P$,  
where $\rho''$ is the union of  $\rho'(CB)$ and  $\rho(BA)$, 
then  $\rho(AB)$ and  $\rho'(CB)$ offer
likely alternatives.  Such configurations of trajectories are referred
to as the interfering alternatives.  The intensity of particles
transmitted to $B$ by the equally likely alternatives must be equal to
the sum of the intensities at $A$ and $C$ associated with the respective
trajectories.  Such a union of paths is 
indistinguishable from a pair of monotonic physical paths since $\kappa(B)$ 
may be adjusted such that $\rho'(CB)$ and $\rho(AB)$ are both physical which 
does not alter the relevant physical content.

{\samepage
\section{Physical Paths}

As a prelude to a more precise treatment of motion in Sec. 5, an approximate
description of a few phenomena is given in this section,
which also clarifies the properties of a multiplicity of physical
trajectories.  

\subsection{Motion of a particle.}  
Consider a physical system described by
a Lagrangian $L(\dot{x},x)$ with $\rho_{\!_S}$ being 
the resulting classical path.  
For convenience, it is assumed that $L$ does not depend on $\tau$ explicitly. 
However, $\tau$-dependence may be included without a significant change in
the following analysis.  For a free particle, $L = L^P$. 
For an undisturbed particle,  the equivalent points on $\rho_{\!_S}$  
are given by
}
\[  S(B_j,B_{j+1},\tau_j,\tau_{j+1}) = 2 \pi  \]
where $S( \hspace{.15in}  )$ denotes Hamilton's principal function.
The action $S_{B'A'}(\rho')$ along a trajectory  $\rho'(A'B')$ 
in a small neighbourhood
of  $\rho(AB)$ is given by
\begin{eqnarray}
 \left( S_{B'A'}(\rho')-S_{BA}(\rho) \right) &=& 
 \int_{\rho(AB)} \delta x^\mu \left[ \frac{ \partial L}{\partial x^\mu}
 - \frac{d}{d \tau} \frac{ \partial L}{ \partial \dot{x}^\mu} \right] d\tau
   \nonumber \\ 
                                               & &
   + \left[ \frac{\partial L}{\partial \dot{x}^\mu} \delta' x^\mu
          - H \delta' \tau \right]_A^B + O(\delta^2)
\end{eqnarray}
by standard methods.  Here $\delta' x^\mu$, $\delta' \tau$  
correspond to the variation of the
end points $A$, $B$ to $A'$, $B'$, and $H$ is the Hamiltonian.  
The term $O(\delta^2)$ is
the integral along  $\rho(AB)$ of an argument, containing functions of 
second or higher
order in $(\delta x)$ and $(\delta \dot{x})$.

If $\rho = \rho_{\!_S}$ , then the first term on the right side of (16) is 
equal to zero.
Hence $S_{B'A'} (\rho') = S_{BA} (\rho_{\!_S})$ for some values of  
$\delta' x = O(\delta^2 )$.  Therefore the
trajectories in a $\delta x$ neighbourhood of a physical 
classical path  $\rho_{\!_S}(B_j B_{j+k})$
are also physical and their end points are confined to $(\delta^2)$
neighbourhoods of $B_j$ and $B_{j+k}$.  Thus the intensity transmitted 
by paths in
a $\delta x$ neighbourhood of a classical trajectory 
is concentrated in $(\delta^2)$
neighbourhoods of the equivalent points on $\rho_{\!_S}$.  
Let $\rho$ be a path
transmitting intensity outside $(\delta^2)$ neighbourhood of $\{ B_j \}$. 
Since $\rho$ is not a solution 
of the Euler-Lagrange
equation, the first term in (16) dominates which is $O(\delta x)$.  
Repeating the
above argument, we have that the intensity transmitted by trajectories
in a $\delta x$ neighbourhood of $\rho$ is spread over a $\delta x$ 
neighbourhood of points
outside $(\delta^2)$ neighbourhood of $\{ B_j \}$.  Further, the magnitude 
of the first
term in (16) increases as $\rho$ is removed farther from the classical
trajectory.  Therefore the contribution to the intensity decreases
accordingly.  Some intensity is also transmitted by the interfering
alternatives whose monotonic segments are non-physical.  In a
homogeneous space, such paths are roughly evenly distributed about the
classical trajectory implying a uniform distribution of the associated
intensity.  The properties of such paths will be described in more
detail in Sec. 4.2 where their impact is greater.

Assuming that the particles originate in a small region about a point $A$,
intensity should be expected to be higher near the points equivalent to
$A$ and to decrease away from them, creating a wave-like pattern over a
uniform background.  On a classical scale, the segments between 
$B_j$ and $B_{j+1}$
are negligibly small.  Also for macroscopic trajectories, the
contribution of the first term in (16) is enormous as one moves away
from a purely classical trajectory, owing to the large interval of
integration.  Therefore, the contribution to the variation of the
intensity over a wavelength, between $B_j$ and $B_{j+1}$ , must 
come from extremely
small neighbourhoods of the long trajectories, and from larger
neighbourhoods of the shorter ones, which are still small on a classical
scale.  Thus on a macroscopic scale, the particles from $A$ to $B$ travel
along narrow beams centered about the classical trajectories.  

\subsection{The double-slit experiment.}  The interfering alternatives play a
prominent role in the double slit experiment.  In this setup, identical
particles are allowed to pass through two slits at $A$ and $A'$, and
collected on a distant screen at a point $B$.  The following treatment is valid
in the presence of a field.  As explained in Sec. 4.1,
the particle paths may be assumed concentrated about the classical
trajectories from $A$ to $B$ and from $A'$ to $B$.  If one of the beams is
blocked, then the intensity observed in a neighbourhood of $B$ should
behave as deduced in Sec. 4.1 for a free particle.  However, if
the intensity is transmitted by both of the beams, then a multitude of
the interfering alternatives is allowed.  Existence of such paths and
their influence on the intensity distribution is studied next.  

In view of the physical equivalence of A and A' and that of the
particles, one has that  $\kappa(A) = \kappa(A')$, 
$\Phi_A^P = \Phi_{A'}^P$, and hence $\Phi_A = \Phi_{A'}$.
However, because of
an interaction with the detecting instrument at $B$, $\kappa(B)$ 
may not be equal to $\kappa(A)$. 
For the interfering alternatives, the value of $\kappa(B)$ is the same for both 
of the monotonic segments (Sec. 3).  Substitutions in (13) show that 
the paths are the solutions of
\begin{equation}
      exp \left[ i \left( \int_{\rho(AB)} dS(x,\tau) - 
             \int_{\rho'(A'B)} dS(x,\tau) \right) \right]\Phi_A = \Phi_A
\end{equation}
For the classical trajectories $\rho = \rho_{\!_S}$  
and $\rho' = \rho'_{\!_S}$,  
(17) is solved by
\[
    \left( S_{BA} (\rho_{\!_S}) - S_{BA'} (\rho'_{\!_S}) \right) = 2 \pi j
\]
where $j$ is an arbitrary integer and the action in this case is
Hamilton's principal function or the arc-length in $\cal M$.  Classical paths
are characterized by a constant velocity $\bar{u}$.  This reduces the solution
to $\Delta r = 2 \pi j/m \bar{u}$, where $\Delta r$ is the difference 
between the path-lengths of
$\rho_{\!_S}(AB)$ and $\rho'_{\!_S}(A'B)$.  Therefore $\rho_{\!_S}(AB)$ and 
$\rho'_{\!_S}(A'B)$ are interfering alternatives
whenever $\Delta r = 2 \pi j/m \bar{u}$.  

Let $B(\varepsilon)$ be the point on the screen such that
\begin{equation}
 \left( (S_{B(\varepsilon)A} (\rho_{\!_S}) - 
         S_{B(\varepsilon)A'} (\rho'_{\!_S}) \right) = 2 \pi (j + \varepsilon)
\end{equation}
for a fixed $j$ and each $0 \leq \varepsilon \leq  1/2$.  
In the following we study the
variation of the intensity as $\varepsilon$ varies in the prescribed 
interval which is
sufficient to describe it on the entire screen.

It follows from the analysis of Sec. 4.1, that 
$S_{CA} (\rho) = S_{B(\varepsilon)A} (\rho_{\!_S})$,
$S_{C'A'} (\rho') = S_{B(\varepsilon)A'} (\rho'_{\!_S})$,
for $\rho$, $\rho'$ in $\delta x$ neighbourhoods of 
$\rho_{\!_S}$, $\rho'_{\!_S}$ 
respectively, where $C$ and $C'$ vary
over a $(\delta^2)$ neighbourhood of $B(\varepsilon)$ on the screen for 
a fixed $\varepsilon$.  Therefore,
by varying the paths over a $(\delta x)$ width of the beam and over a 
$(\delta^2)$
neighbourhood of $B(\varepsilon)$ it is possible to satisfy
\[
  \left( S_{DA} (\rho) - S_{DA'}(\rho') \right) = 2 \pi (j + \varepsilon)
\] 
for most of the paths.  In fact cancellations favour this equality which
can be easily seen, in particular for the cases when $\rho_{\!_S}$, 
$\rho'_{\!_S}$ are extremals as
is presently the case.  This conclusion is valid for other points in the
vicinity of $A$ and $A'$ also.  For $\varepsilon = 0$, this implies that 
there is a large
concentration of interfering alternatives reaching about $B(0)$ and hence
the intensity in a $(\delta^2)$ neighbourhood of $B(0)$ is almost equal to the
intensity in $\delta x$ neighbourhoods of $\rho_{\!_S}(AB(0))$ and  
$\rho'_{\!_S}(A'B(0))$.  For $\varepsilon \neq 0$, the
configuration of the paths $\rho_{\!_S}(AB(\varepsilon))$ and 
$\rho'_{\!_S}(A'B(\varepsilon))$ is obviously non-physical. From the above
argument, a large number of paths in $\delta x$ neighbourhoods of 
$\rho_{\!_S}(AB(\varepsilon))$ and $\rho'_{\!_S}(A'B(\varepsilon))$ 
are excluded from
combining to form the interfering alternatives and hence unable to
transmit the intensity in a $(\delta^2)$ neighbourhood of 
$B(\varepsilon)$.  Still there are many paths capable of transmitting intensity
about $B(\varepsilon)$ for $\varepsilon \neq 0$, which are described below. 

It follows from (16) that for trajectories $\rho(AB(\varepsilon))$,  
$\rho'(A'B(\varepsilon))$ in $\delta x$
neighbourhoods of $\rho_{\!_S}(AB(\varepsilon))$, 
$\rho'_{\!_S}(A'B(\varepsilon))$ respectively,
\[
  (S_{B(\varepsilon)A}(\rho) - S_{B(\varepsilon)A}(\rho_{\!_S})) 
  = O(\delta^2)
\]
and
\[
  (S_{B(\varepsilon)A'}(\rho') - S_{B(\varepsilon)A'}(\rho'_{\!_S})) 
  = O(\delta^2)
\]
We have used the fact that the first term on the right side of (16) is
zero as the curves are varied about the classical trajectories and the
second term is zero as the end points are kept fixed.  For these curves,
we have
\begin{equation}
    (S_{B(\varepsilon)A}(\rho) - S_{B(\varepsilon)A'}(\rho')) 
  = 2 \pi (j + \varepsilon) + O(\delta^2)
\end{equation}
Since there are distortions for which $O(\delta^2)$ term is non-zero and its
magnitude is large in natural units, it is possible to adjust the curves
$\rho$, $\rho'$   such that
\begin{equation}
   (S_{B(\varepsilon)A}(\rho) - S_{B(\varepsilon)A'}(\rho')) = 2 \pi k
\end{equation}
with $k = j$ or $(j+1)$, most likely $j$.  This implies that 
$\rho(AB(\varepsilon))$ and
$\rho'(A'B(\varepsilon))$ form a pair of interfering alternatives.  
Since $\rho(AB(\varepsilon))$, $\rho'(A'B(\varepsilon))$
are non-classical trajectories, it follows as in Sec. 4.1 that while
there is a multitude of paths satisfying (20), in $\delta x$ neighbourhoods of
the central paths, their end points are spread over a $\delta x(\varepsilon)$
neighbourhood
of $B(\varepsilon)$.  This implies that the amount of intensity that is 
concentrated
in a $(\delta^2)$ neighbourhood of $B(0)$ is spread over a 
$\delta x(\varepsilon)$ neighbourhood of $B(\varepsilon)$.
Consequently, a rapid decrease in the intensity is expected as $\varepsilon$
increases away from zero.

As $\varepsilon$ increases further, it is seen from (19) that the 
neighbourhood $\delta x$ must
be increased to satisfy (20), i.e. $\rho(AB(\varepsilon))$, 
$\rho'(A'B(\varepsilon))$ must be moved farther
away from the solutions of the Euler-Lagrange equations.  Thus the
magnitude of the first term on the right side of (16) integrated along
$\rho(AB(\varepsilon))$, $\rho'(A'B(\varepsilon))$ increases as $\varepsilon$
increases for each fixed variation $\delta x$.  As
above, $O(\delta x(\varepsilon))$ increases with $\varepsilon$, 
implying a decrease in the intensity.

The above arguments also imply a symmetric intensity distribution as 
$\varepsilon$ is
varied over the interval zero to -1/2, and a repeat of the pattern as $j$
is varied over the integers.  Thus an interference pattern should be
observed on the screen over a background of almost uniform but
relatively low intensity as the major contributions have been estimated here.

Similar arguments may be used to estimate the variations in the
intensity about peaks as $j$ varies, resulting in a decrease in the
intensity as $\mid j \mid$ increases.  
This result is based on the fact that the
term $O(\delta^2)$ for each $j$, may be expressed as a sum of two terms, 
one being
$j$-independent and the other, directly proportional to $\mid j \mid$.

Availability of two interfering beams originating at $A$, $A'$ and the
equivalence of the physical conditions at these points have played a
crucial role in the above analysis.  As explained before, if one of the
beams is blocked, the interference pattern is destroyed.  Also, such a
distribution should not be expected to result if the equivalence of $A$
and $A'$ is violated.  This situation arises when an attempt is made to 
observe the particle anywhere along the trajectory.  Interaction with the
detecting instrument changes the classical momentum of the particle 
say by $\Delta P$.  It is straight forward to estimate the 
change $\Delta S$ in the action which is very large for the macroscopic
trajectories.  This enables one to estimate $\kappa$. Consequently, 
a point B that
was physical previously, either is no longer so or if physical, corresponds
to a large value of $\mid j \mid$.  In either case, the intensity transmitted
to B by the interfering alternatives is negligible.  Hence, the two beams
transmit intensity as the classical beams of particles.

Above considerations indicate a wave-like behaviour of microscopic
particles observed macroscopically as a collection while behaving as
particles individually.  This is in agreement with the observed
behaviour [1, pp. 2-5].  These results obtained here from (13), are
known to inspire the formalism of quantum mechanics.

\subsection{The Aharonov-Bohm effect.}   Additional insight into the behaviour
of the particles as implied by the present extension may be gained by
considering their response to a non-zero gauge field, as follows.  The
gauge transformation obtained by replacing $\phi_\mu$ by $\hat{\phi}_\mu$
will be denoted by $\hat{U}_{BA}(\rho)$. 
Let $\{ \rho \}$, $\{ \hat{\rho} \}$ be the collections of the solutions 
of (13), with $\phi_\mu$, $\hat{\phi}_\mu$ respectively.  
Assume that $U_{BA}(\rho) \neq \hat{U}_{BA}(\rho)$ for a
solution $\rho(AB)$.  If $U_{BA} (\rho)$ is replaced by $\hat{U}_{BA}(\rho)$
in (13), then $\rho(AB)$ is no
longer a solution.  The same conclusion holds for a path $\hat{\rho}(A'B')$.
Thus,
if the inequality holds for some of the solutions of (13) with 
$\phi_\mu$, or with $\hat{\phi}_\mu$, then the collections $\{ \rho \}$,
and $\{ \hat{\rho} \}$ 
of the physical paths are not identical. 
Therefore a change of potentials from $\phi_\mu$ to $\hat{\phi}_\mu$ 
should in general produce an
observable effect.  However, if $U_{BA}(\rho') = \hat{U}_{BA}(\rho')$ 
for each $\rho'(AB)$ in a
collection $\{ \rho' \}$ large enough to include the union of 
$\{ \rho \}$ and $\{ \hat{\rho} \}$, then (13)
remains the same equation under the change from $\phi_\mu$ to 
$\hat{\phi}_\mu$.  Consequently, a
change of potential from $\phi_\mu$ to 
$\hat{\phi}_\mu$  would not change the solutions $\{ \rho \}$.  Since
the set of physical paths remains the same under this change, the
response of the particles must remain unchanged also.  Therefore, such a
change of potentials will not alter the outcome of an experimental observation.

As an application, consider the Aharonov-Bohm effect [7].  In the
corresponding experimental set up, the electrons travel in beams centered 
about paths $\rho(ACB)$ and $\rho'(ADB)$, enclosing a non-zero magnetic
field but shielded from it.  Chambers used reflectors at $C$ and $D$ to
obtain a configuration of piece-wise classical narrow beams centered
about $\rho(AC)$, $\rho(CB)$, $\rho'(AD)$ and $\rho'(DB)$ [8].  
The magnetic field was generated
by placing a long coil carrying an electric current between the
reflectors and perpendicular to the plane of the beams with one end in
the plane.  The electron beams were further shielded from the magnetic
field.  As the current in the coil is varied, the magnetic field varies
accordingly.  The classical Lagrangian for this system is the same as
for the Lorentz equation.

As in the case of the double slit experiment, most of the electrons are
transmitted by the interfering alternatives with parameter value
increasing from $A$ to $B$ along $\rho(ACB)$ and decreasing from $B$ 
to $A$ along
$\rho'(BDA)$, taking value $\tau_{\!_B}$ at $B$.  The estimates obtained 
in the treatment of
the double slit experiment are valid for the present case as they were
not restricted to a free particle.  Some consideration should be given to
the reflectors at $C$ and $D$.  Because of the continuity of the physical
paths at points about $A$, $B$, $C$, and $D$, 
$\kappa( \hspace{.15in} )$ cancels out.  
From Sec. 4.1, we
have that most of the intensity transmitted along $\rho(AC)$ reaches a small
neighbourhood of $C$ which remains almost within a macroscopically narrow
beam.  By the same argument, most of this intensity reaches a small
neighbourhood of $B$.  The same comment is valid for $\rho'(ADB)$.  
The intensity
along both of the beams is assumed equal.  Consequently, the arguments
of Sec. 4.2 can be used to conclude the existence of a similar
interference pattern on the screen.

It follows from (13) that the interfering alternatives for an
electro-magnetic potential $\phi_\mu$ are the solutions of:
\begin{equation}
 \exp \left[ i \oint (dS^P(x,\tau) - \phi_\mu dx^\mu) \right] \Phi_A = \Phi_A
\end{equation}
where $S^P(x,\tau)$ is the free particle part of the action and
the integration is along the closed curves $\rho_{\!_c}(ACBDA)$.  
Here the group element $U_{BA}(\rho)$ is given by
\[
 U_{BA}(\rho) = \exp ( i \int_{\rho(AB)} \phi_\mu dx^\mu )
\]
It is clear that $\phi_\mu$-dependent part in (21) is $U_{ABA}(\rho_{\!_c})$
which is given by $U_{ABA}(\rho_{\!_c}) = \exp (i F(\phi))$
where $F(\phi)$ is the magnetic flux enclosed by $\rho_{\!_c}$.  
As $\rho_{\!_c}$ is
distorted, $F(\phi)$ remains unchanged as long as the distorted closed path
encloses the flux, which covers all of the paths of significance here
as all of them surround the coil.

As $F(\phi)$ varies to $F(\hat{\phi})$, 
$U_{ABA}(\rho_{\!_c}) \neq  \hat{U}_{ABA}(\rho_{\!_c})$ for any
$\rho_{\!_c}$  unless
\begin{eqnarray}
(F(\phi) - F(\hat{\phi})) & = & \oint (\phi_\mu 
                               - \hat{\phi}_\mu)dx^\mu \nonumber \\
                          & = & 2 \pi j
\end{eqnarray}
with an arbitrary integer $j$.  Whenever (22) is satisfied, 
$U_{ABA}(\rho_{\!_c}) = \hat{U}_{ABA}(\rho_{\!_c})$ for
each curve $\rho_{\!_c}$ and hence the experimental observation with 
$\hat{\phi}_\mu$ must be the same
as with $\phi_\mu$.  Thus the interference pattern on the screen should repeat
itself periodically as the potential is varied continuously.  The period
is defined by (22).

Let $\phi_\mu(\varepsilon)$ be a one parameter family of potentials with 
$0 \leq \varepsilon \leq 1$, such that
$(F(\phi(1)) - F(\phi(0))) = 2 \pi$ , i.e., $\varepsilon$ 
covers one period.  The intensity
patterns corresponding to $\phi_\mu(0)$ and $\phi_\mu(1)$ are 
indistinguishable.  Let the
solutions of (21) with $\phi_\mu$ replaced by $\phi_\mu(\varepsilon)$
be $\{ \rho(\varepsilon) \}$.  
Owing to the continuity
of $F(\phi(\varepsilon))$ with respect to $\varepsilon$, 
$\{ \rho(\varepsilon) \}$ should vary continuously, implying a
continuous variation of the corresponding interference pattern.  As 
$\varepsilon$ approaches one, the distribution of the intensity 
must return to the
same as for $\varepsilon = 0$.  Thus, each interference fringe 
should be expected to
shift as $\varepsilon$ varies from zero to one, from its position to 
the original location of the next.

Above conclusion agrees with the experimental observation [8,9].  It
is pertinent to remark that the indistinguishability of 
$\phi_\mu$ and $\hat{\phi}_\mu$ that satisfy
(22) is a direct consequence of (21) which is obtained from (13) and the
fact that the physical paths in this case are closed in $\cal M'$.  
For this part
of the conclusion, no estimates are needed.  

The Aharanov-Bohm effect is an implication of the quantum mechanical
equations [7] which were developed from different premises than the
present formalism.  Ingredients of the quantum mechanical deduction of
this effect are the representation of 
the momenta $p_\mu$ by $-i \partial_\mu$ and the
corresponding extension of the classical coupling scheme 
$( p_\mu - \phi_\mu)$.  The
former was inspired by the observed wave-like behaviour of particles and
the later, in addition to being intuitive, sets $\alpha= i$ in the 
London-Weyl [4,10]
description of electro-magnetism.  Here the major aspects of the
Aharonov-Bohm effect are deduced directly from (13) without an appeal to
any other theory. 

Above considerations show that the wave-like behaviour of a
particle in motion is a result of the properties of the physical
paths. However, there is a crucial difference as described below.
Consider the double slit experiment. If the intensity pattern on the
screen is a result of a wave motion, then there must be a point of
zero intensity in between two bright regions. According to the present
formulation, a point of minimum intensity exists but it can be seen
that there must be some physical paths reaching every point on the
screen, resulting in some intensity everywhere. If accurate enough
determination of the intensity can be made, it may be possible to test
whether the present theory or Quantum Mechanics provides a better
description of motion. Nevertheless, major contribution to the
intensity in the present formulation is the same as predicted by the
wave motion. Thus one may use the results from the wave theory in
building a theory of mechanics, at least approximately. While the
above considerations justify use of the results from the theory of
waves, it should be remarked that it is only for convenience rather
than a physical attribute of the particles.

\section{Equation of Motion}

The classical action principle assigns a unique trajectory to a particle
in motion between two points.  The present extension (13), on the other
hand, assigns many paths, but not all curves are allowed.  Since it is
impossible to assign a unique trajectory to a particle, as an
alternative, one may describe its motion in terms of the intensity of
the particles transmitted to a region in $\cal M$ or $\cal M'$ 
by the physical trajectories.  This was done in Sec.\ 4 for
a beam of free particles and for the double-slit experiment, but only
approximately.  Approximations were made in obtaining the estimates and
by retaining only the major contributions.  In a complete theory, all
physical paths must be included and the contributions must be computed exactly.
While such a theory is possible, it will require quite intricate
computations for which a machinery is not yet developed. An
approximate theory may be developed by exploiting the wave-like
behaviour of the particles deduced in Sec. 4. In addition to
simplifying the manipulations, this relates the present formulation
with Quantum Mechanics which is instructive in itself.

Wave-like behaviour of particles and a possibility of describing their
motion in terms of the probability densities associated with a
collection of trajectories led Feynman to develop his path integral
formulation of non-relativistic quantum mechanics [1,11].  The
formalism was extended to the relativistic case in an analogous manner 
by introducing a proper
time-like evolution parameter [12].  The wave-like behaviour of the
particles was used to conclude that the intensity is the absolute square
of the amplitude obtained by the law of superposition.  The amplitude
associated with a path $\rho(AB)$ was taken to be proportional  
to $\exp ( i S_{BA}(\rho))$ 
which was based on a deduction by Dirac [13] of the behaviour of a
quantum mechanical particle.  Present formulation associates a phase-factor
equal to $\exp ( i S_{BA}(\rho))$ with $\rho(AB)$ whenever a 
classical description is
possible in terms of a Lagrangian.  
The phases associated with a multiplicity of paths are shown in Sec. 4 to
interfere in a manner that imparts wave-like properties to the
particles in motion.  A precise determination of a multitude of physical
trajectories follows from (13).  Thus all of the necessary assumptions
required for the formulation of Feynman's postulates have been deduced
from (13).  
It is straight forward to check that the assumption of particle following 
any out of all possible paths is extraneous to Feynman's postulates.
Having yielded its basic assumptions, the gauge mechanical
principle finds a natural expression within the framework of the path
integral formalism.  However, only the physical paths should be included
in the computation of the total contribution.\\[2ex]
{\em Postulate 1}.~~~The probability of finding a particle in
a region of space-time is the absolute value of the
sum of contributions from each physical path or its segment in the region.
\\[1.5ex]
{\em Postulate 2}.~~~The contribution at a point $C$ of a physical path
$\rho(AB)$ is equal to $K V_{CA}'(\rho) \Phi_A$ 
where $K$ is a path-independent constant.  \\[.6ex]

Since the assumptions underlying the above postulates are deduced from
(13), the formalism is self-consistent and based essentially on one
assumption.  Postulate 2. provides a mechanism for a
computation of the total contribution from all trajectories by the
techniques developed originally for the path-integral formulation.  An
equation of motion is developed below by this procedure and by isolating
the contribution of the physical paths.  Postulate 1. provides a means
to obtain experimentally observable quantities from the solutions of the
equation of motion.

Consider a point $C$ on a physical path $\rho(A'B')$. Let $\rho(AB)$ be
the shortest segment of $\rho(A'B')$ containing $C$ such that $A$ and
$B$ are equivalent to $A'$ and $B'$ respectively. Consider the pair of
points $A$ and $A'$. The pair $B$, $B'$ is treated similarly. In view of
the equivalence, $V_{AA'}(\rho)\Phi_{A'}^P = \Phi_{A'}^P = \Phi_{A}^P$, 
we have that 
$V_{CA'}(\rho)\Phi_{A'}^P = V_{CA}(\rho)\Phi_{A}^P$. Thus
the contribution from $\rho(A'C)$ is the same as that from $\rho(AC)$.
Therefore it is sufficient to consider the minimal curves $\rho(AB)$
instead of any larger physical paths containing $\rho(AB)$. 
As indicated in Sec. 3, interfering alternatives are included in this 
treatment.

The next step is to parameterize the minimal physical paths in a way
that enables one to isolate their contribution.
Since a single parameter is needed for all of the curves, standard
parameterization by arc-length is inadequate. A suitable parameter was
found in ref.[6] as follows. Let $u'_\mu = \sum u_\mu$ where
$\sum$ denotes the sum over all
paths of the type $\rho(AB)$ with $A$ being a variable point. For any
such collection of curves, there is a 
Lorentz frame $\cal L$ in which $u_\mu' = 0$ for $\mu=1,2,3$. A particle
may thus be treated as being located at the origin of $\cal L$. Incidentally,
the origin of $\cal L$ coincides with the centre of mass of a fluid of
uniform density and total mass $m$ with an infinitesimal element
flowing along each of $\rho(AB)$ and with the arc-length in an appropriate 
Finsler space [3, Ch.\ 3.2]. Let $z(\tau)$ be a
parameterization of each path $\rho(AB)$ with $z(0)=A$, where $\tau$ is
the proper time of $\cal L$. In $\cal L$, each of the curves $\rho(AB)$
coincides with the straight line along $\tau$. Therefore, 
$V_{CA}(\rho) = \exp(im\tau)$ and hence $B = z(2\pi/m)$. From Postulate
2, the contribution $\psi'(x,\tau)$ at $C = x$ is given by 
\begin{equation}
  \psi'(x,\tau) = \sum K' V'[x,z(\tau)] \Phi[z(0)]
\end{equation}
where the sum is over all paths passing through $x$ at $\tau$; 
$\Phi [z(0)] = \Phi_A$ and for each
$z(\tau)$, $V'[x,z(\tau)] = V_{CA}'[z(\tau)]$.
The sum is the limit of a finite one with constant $K'$ depending on the
number of terms. Because of the continuity of the paths, 
the number of curves for
$\tau=0$ is the same as for $\tau=2\pi/m$. Also, for each physical path
$z(\tau)$, $V[x,z(0)] = V[x,z(2\pi/m)] = 1$, i.e., 
$V'[x,z(0)] = V'[x,z(2\pi/m)] = \kappa^{-1}(C) \kappa(A)$. It follows
that 
\begin{equation}
   \psi'(x,0) = \psi'(x,2\pi/m)
\end{equation}

The boundary condition given by (24) provides a means to retain the
contribution in (23) from the physical paths. Thus the proper time $\tau$ of
$\cal L$ acquires a physical significance, which
is treated below as an independent parameter as in [14]. The
following derivation is essentially the same as in the standard path
integral formulation.

Let $[0,2\pi/m]$ be divided into $N$ equal intervals $[\tau_j,\tau_{j+1}]$,
$j=0,1,...,\mbox{$N-1$}$; with $\tau_0 = 0$, and $\tau_N = 2\pi/m$. 
Consider all
of the paths with $z(\tau_k) = {(x)}_k$. By the standard argument 
the function $\psi'[{(x)}_k,\tau_k]$, for each $k$, is given by
\pagebreak[1]
{\samepage
\begin{eqnarray}
  \psi'[(x)_k,\tau_k] &=& 
    \int U^P[{(x)}_0,{(x)}_1] \cdot\cdot\cdot U^P[{(x)}_{k-1},{(x)}_k] 
    \nonumber \\
 && ~~~~~\times ~ U[{(x)}_k,{(x)}_{k-1}] \cdot\cdot\cdot U[{(x)}_1,{(x)}_0]
         \Phi[z(0)]                                              
    \nonumber  \\
 && ~~~~~\times ~ \frac{d{(x)}_0}{Q} \cdot\cdot\cdot \frac{d{(x)}_{k-1}}{Q}
\end{eqnarray}
}
where~~~
$U^P[{(x)}_{j+1},{(x)}_j]  ~~=~~  { \{ U^P[{(x)}_j,{(x)}_{j+1}] \} }^{-1}
  ~~=~~  U_{B'A'}^P[z(\tau)]$, \\*
$U[{(x)}_{j+1},{(x)}_j] = U_{B'A'}[z(\tau)]$ with $A'={(x)}_j$, 
$B'={(x)}_{j+1}$, and $Q$ is a normalization constant. Set ${(x)}_k=y$,
$\tau_k = \tau$, ${(x)}_{k+1} = x$ and $\tau_{k+1} = \tau_k + \epsilon$. 
It follows from (25) that
\begin{equation}
  \psi'(x,\tau + \epsilon) = \frac{1}{Q} 
                             \int U^P(y,x) U(x,y) \psi'(y,\tau) dy
\end{equation}

A curve $z(\tau)$ in $\cal M$ may be arbitrarily closely approximated by
$z_N(\tau)$ for large enough $N$, where $z_N(\tau_j)=z(\tau_j)$,
$j=0,1,...,N$; and in each of the intervals $[\tau_j,\tau_{j+1}]$,
$z_N(\tau)$ is the geodesic line. The element  
$U^P(y,x) = U_{y,x}^P[z(\tau)]$ may be approximated by 
\[
    U_{y,x}^P[z_N(\tau)] = \exp \left[ \; iS^P(x,y) \; \right]
\]
where $S^P(x,y)$ is Hamilton's principal function for a
`free' particle of mass $m$ from $x$ to a variable point $y$.
Here the
Lagrangian is $L^P$ with $\tau$ being the proper time of $\cal L$. The
action is given by 
\[
  S^P(x,y) = - \frac{m}{2\epsilon} g_{\mu\nu} \xi^\mu \xi^\nu 
           - \frac{m}{2} \epsilon
\]
where $\xi^\mu = (x^\mu - y^\mu)$. Also, $U(x,y)$ is approximated by 
$U_{x,y}[z_N(\tau)]$ up to the desired order which is given by 
\begin{eqnarray*}
   U_{x,y}[z_N(\tau)] &=& 1 + i \phi_\mu(x) \xi^\mu                     
    - \frac{1}{2} \left[ i\phi_{\mu,\nu} + \phi_\mu \phi_\nu \right] 
      \xi^\mu \xi^\nu                            \\
   && ~~ + ~{\rm higher~order~terms}.
\end{eqnarray*}
Let $\psi(x,\tau) = \exp(im\tau/2)\psi'(x,\tau)$, then it follows from
(24) that 
\begin{equation}
   \psi(x,0) = - \psi(x,2\pi/m),
\end{equation}
With the above substitutions, from (26), we have
\begin{equation}
  \psi(x,\tau+\varepsilon) = \frac{1}{Q} \int \exp
    \left[ -\frac{im}{2\epsilon} g_{\mu\nu} \xi^\mu \xi^\nu \right]
    U_{x,y}[z_N(\tau)] \psi(x-\xi,\tau) d\xi
\end{equation}
Eq. (28) holds exactly in the limit of infinite $N$, equivalently
$\epsilon=0$. As such it holds up to the first order in $\epsilon$,
which is sufficient for the present.

Expanding $\psi(x,\tau+\epsilon)$ and $\psi(x-\xi,\tau)$ in a Taylor
series about the point $(x,\tau)$ and comparing the coefficients of
$\epsilon^j$, $j=0,1$, yields $Q=-i(2\pi\epsilon/m)^2$ and 
\begin{equation}
   i \frac{\partial \psi}{\partial \tau} 
 = - \frac{1}{2m} \Pi_\mu \Pi^\mu \psi
\end{equation}
where $\Pi_\mu = (i \partial/\partial x^\mu \cdot 1 + \phi_\mu)$. In
view of the boundary condition (27), $\psi$ may be expressed as
\[
   \psi(x,\tau) = \sum_{-\infty}^{\infty} \psi_n(x) \omega_n(\tau)
\]
where for each $n$, 
$\omega_n(\tau) = \sqrt{m/2\pi} \; \exp[i(n+1/2)m\tau]$ and $\psi_n$ satisfies 
\begin{eqnarray}
\Pi_\mu \Pi^\mu \psi_n & = & (2n+1)m^2 \psi_n  \nonumber \\*  
  &  &  n=0,\pm 1,\pm 2, \cdot\cdot\cdot 
\end{eqnarray}
For $n=0$, (30) reduces to the Klein-Gordon
equation in an electro-magnetic field.        

Equation of motion (29) termed the generalized 
Schr\"{o}dinger equation, was first conjectured by
St\"{u}ckelberg [15]. The boundary condition (27) is a direct
result of the definition of the physical paths provided by (13).
As shown above, this boundary condition is crucial
in relating (29) to the Klein-Gordon equation. If all trajectories are
allowed to contribute, the resulting equation is still (29) but without
the boundary condition (27). Feynman [12] used this equation 
to deduce the Klein-Gordon equation by restricting the solution to the
form $\psi_0(x) \omega_0(\tau)$.
Present treatment relates (29) with the Klein-Gordon equation (30) quite
naturally. Further to the arguments of Sec. 4, this result provides
additional support for the assumption (13).

\section{Further Developments}

The above procedure has also been used to develop an equation of
motion in a Riemannian space where the resulting theory is
conceptually clearer [16]. In particular, the arc-length serves as an
appropriate evolution parameter which also indicates that a more
accurate theory would be easier to develop in the setting of a
Riemannian space. For the present, the analogue of the generalized
Schr\"{o}dinger equation in a Riemannian space reads as 
\begin{equation}
   2im' \frac{\partial\psi}{\partial\tau}  
   = [ \partial_\mu \partial^\mu + \frac{1}{3} R ] \psi
\end{equation}
where $\partial_\mu$ are the components of the covariant derivative, R is
the curvature scalar and $\mu$ runs over the dimension of the space. 
The parameter $m'$ is determined by the classical Hamilton's equations.  
For a gravitational field $m'=m$.
The boundary condition \( \psi(x,2\pi /m') = - \psi(x,0) \) still holds which
reduces (31) into infinitely many equations:
\begin{equation}
  - \partial_\mu \partial^\mu \psi_n  = [ (2n+1)m^2 + \frac{1}{3}R ]
  \psi_n \hspace{.1in} ,
    \hspace {.25in} n = 0, \pm 1, \pm 2, ....
\end{equation}
For n=0, in standard units (32) reads as
\begin{equation}
  -\hbar^2 \partial_\mu \partial^\mu \psi_0  
  = [ m^2  c^4   +  \frac{1}{3} \hbar^2   R ] \psi_0
\end{equation}
where $c$ is the speed of light, $\hbar=h/2 \pi$ and $h$ is Planck's constant.

Motion of a charged particle in an electro-magnetic field may be
described in the setting of a Riemannian space in the Kaluza-Klein
framework [17]. The equations of motion may be obtained as special
cases of the equations in the Riemannian spaces or independently [18].
The resulting generalized Schr\"{o}dinger equation is given by
%
%
\begin{eqnarray}
\frac{\partial \psi}{\partial \tau} = 
\frac{1}{2im'}\left[
\left(
\frac{\partial }{\partial x^\mu}-\phi_\mu  \frac{\partial }{\partial x^5}
\right)
\left(
\frac{\partial }{\partial x_\mu}-\phi^\mu  \frac{\partial }{\partial x^5}
\right)
- \frac{1}{12}F_{\mu\nu}F^{\mu\nu} - (\frac{\partial}{\partial x^5})^2 
\right]\psi   \nonumber  \\*[0.5ex]
\end{eqnarray}
where $m'=\sqrt{m^2-e^2}$ with $e$ being the charge.
In view of the compactness of the fifth dimension and the associated
periodicity, $\psi$   may be expanded in a Fourier series:
\[
 \psi = \sum_{k=-\infty}^\infty \psi_k(x,\tau) \exp[iekx^5]
\]
     where $e = p_5$, and since it will cause no confusion, $x$ now
     denotes a point in the Minkowski manifold instead of the Kaluza-Klein.
     Substitution of the expansion for $\psi$ in (34) decomposes it into a set
     of generalized Schr\"{o}dinger type equations with charge quantized in
     units of $e$:
\begin{eqnarray}
-2im'\frac{\partial \psi_k}{\partial \tau}  & = & 
\left[\left( i\frac{\partial }{\partial x^\mu} + ek\phi_\mu \right)
\left( i\frac{\partial }{\partial x_\mu} + ek\phi^\mu \right)
 + \frac{1}{12}F_{\mu\nu}F^{\mu\nu} - (ek)^2 \right]\psi_k  \nonumber \\*[1ex]
 &  & k  =   0,  \pm 1,  \pm 2, ...
\end{eqnarray}
     Further, in view of the boundary condition   
     $\psi(x,2\pi/m')=-\psi(x,0)$ with
     respect to $\tau$,  $\psi_k$  may be expressed as 
\[
\psi_k(x,\tau) = \sum_{n=-\infty}^{\infty} \psi_{kn}(x) \exp[i(n+1/2)m'\tau]
\]
     reducing (35) to
\begin{eqnarray}
\lefteqn{ \left( i\frac{\partial }{\partial x^\mu} + ek\phi_\mu \right) 
 \left( i\frac{\partial }{\partial x_\mu} + ek\phi^\mu \right)\psi_{kn}  = }
   \nonumber \\*
 & & {[(2n+1)(m^2-e^2)+e^2 k^2-\frac{1}{12}F_{\mu\nu}F^{\mu\nu} ]} 
      \psi_{kn} \;\;
   \nonumber \\*[1ex]
 & &  k,n  =  0,  \pm 1,  \pm 2, ....
\end{eqnarray}
     For $n=0$ and $k=1$, (36) is the Klein-Gordon equation with 
     $m^2$  modified by
     $(-F_{\mu\nu}F^{\mu\nu}/12)$, one third of the curvature scalar 
     of the five
     dimensional Kaluza-Klein space. In standard units, the equation for
     $\psi_{10}$  is expressed as 
\begin{equation}
 \left( i\hbar\frac{\partial }{\partial x^\mu} + \frac{e}{c}\phi_\mu \right) 
 \left( i\hbar\frac{\partial }{\partial x_\mu} + \frac{e}{c}\phi^\mu \right)
 \psi_{10}
=  [ m^2 c^4 - \frac{1}{6}G \hbar^2 F_{\mu\nu}F^{\mu\nu} ]\psi_{10}
\end{equation}
     where $G = 6.66\times 10^{-8} {\rm dyn.cm^2 /gm^2}$
     is the universal gravitational constant. 

The above methods are applicable also to the case of a general
gauge field in the setting of the Minkowski manifold [5] or a
Riemannian space in the Kaluza-Klein framework, and to the treatment
of the spinors [19].

The next major step in constructing a complete theory of mechanics in
the present framework would be to abandon the path-integral formalism
and compute the intensity transmitted by the physical paths directly by 
solving the functional equations. 
Comparisons with other theories e.g. Bohmian mechanics is desirable.
The studies of other phenomena e.g. tunneling and behaviour of 
the correlated particles, even with the level of accuracy of Sec. 4,
should prove instructive.
Also, the physical implications of the additional equations arising here should 
be investigated.

\section{Concluding Remarks}

The action principle determines a particle trajectory by requiring
the action to be stationary under all small deformations. In group
theoretical terms, this results in a requirement of equivalence between
the elements associated with a subset of the closed curves up to the
first order only. In this article, the classical action principle is
extended to require the equivalence of the global elements associated with
all of the curves. Solutions of the resulting equation
form an infinite subset, termed the physical paths, to assign to a
particle in motion.

Properties of the physical paths impart wave-like properties to a
particle in motion.
The wave-like behaviour of particles and the
multiplicity of allowed paths form the basis of the path integral
formulation. An imaginary value of $\alpha$ yielded by the present
extension, implies the compactness of the gauge groups which is 
inherent in quantum mechanical equations in gauge fields.
Consequent description of the influence of the field enclosed by a closed
curve on the particles, as is the case with the Aharonov-Bohm effect, is
described by (13) to a large extent without an appeal to any other
theory. Thus the present formulation develops a coherent theory
unifying various treatments underlying the existing quantum mechanics
without involving its usual assumptions. 

The above results lead naturally to Feynman's path integral 
formalism with physical paths being the contributing members.
The criterion imposed by (13)
on the physical paths plays a crucial role in the
deduction of the above results, some of which have been used to justify
the use of the path integral formalism. Thus the present formulation is
self-consistent.

In the present paper we have used a proper time-like parameter to
convert the problem of isolating the contribution from the physical
paths into a boundary condition on (29). This type of parameter was
introduced in a rather {\sl ad hoc} manner by several authors [12,14,15].
Here this parameter 
gains a clearer physical significance. A need for a five-dimensional
relativistic wave equation has been felt for a long time, for the
existing equations suffer from some conceptual difficulties. In response
to this need, St\"{u}ckelberg originally conjectured the generalized
Schr\"{o}dinger equation for a particle in an Abelian gauge field [15].
There is a renewed interest in this equation to interpret it in a
more satisfactory framework than a conjecture, as well as to study its
implications (see e.g., [20]). Present formalism provides a
systematic derivation of the generalized Schr\"{o}dinger equation. 

In addition to accepting the conjecture of St\"{u}ckelberg, 
Feynman selected a particular set of periodic solutions to
deduce the Klein-Gordon equation from the generalized Schr\"{o}dinger
equation. As pointed out above, 
the physical paths are characterized by a boundary condition on (29). 
This boundary condition confines the solution to a set described
by a class of periodic functions. As a consequence, the equation decomposes 
into countably many four dimensional equations, one of them
being the Klein-Gordon equation. Thus the resulting boundary condition
provides an additional justification for the present treatment.

Classical description of motion is quite accurate at the macroscopic scale.  
Quantum Mechanics modifies these results only slightly but conceptually it is 
fundamentally different.  It also appeals to experimental observations for its 
underlying assumptions without offering conceptual clarity.  
The present formulation extends Classical Mechanics yielding these assumptions 
and various conjectures in a coherent framework.  Thus the gauge mechanical 
principle offers a more satisfactory basis for the formulation of mechanics.
In particular, it eliminates the need for a direct assumption of wave nature of
a particle in motion which underlies the well known difficulties with Quantum
Mechanics. It is pertinent to remark that while the present theory associates a 
somewhat objective meaning to a particle in motion, an element of randomness 
remains in the availability of the equally likely, infinitely many paths.

Quantum Mechanics results as an approximation to the present theory, presumably
quite accurate.  Deviations from Quantum Mechanics are pointed out, and
directions for further investigations, and to construct a more accurate and
complete theory, are indicated.

\begin{center}
{\bf Acknowledgements}\\*
\end{center}
The author is thankful to Patrick A. O'Connor for helpful discussions,
encouragement and substantial logistical support without which this
presentation would not have been possible. 


%
%
\end{document}